\documentclass[11pt,twoside]{article}
\usepackage{asp2010}

\newcommand{\SII}{[S\,{\sc ii}]}

\newcommand{\OIII}{[O\,{\sc iii}]}

\resetcounters

\begin{document}

\title{A multi-wavelength investigation of newly discovered planetary nebulae in the Large Magellanic Cloud: Central stars}
\author{Warren Reid $^1$$^,$$^2$
\affil{$^1$Department of Physics and Astronomy, Macquarie University, North Ryde, Sydney, NSW 2109 Australia}
\affil{$^2$Centre for Astronomy, Astrophysics and Astrophotonics, Macquarie University, North Ryde, Sydney, NSW 2109 Australia}
}

\begin{abstract}
Having completed my search for faint PNe in the LMC, including the outer 64 deg$^{2}$ area not covered in the original UKST survey, I now have the most complete number of PNe within any galaxy with which to assess stellar parameters. I present preliminary estimates for planetary nebula central star temperatures for 688 LMC PNe using the excitation class parameter derived from emission lines in the nebula. These are then compared to a photoionisation model in order to evaluate the contribution of metallicity when determining stellar temperatures using only emission lines. I include measurements from my latest confirmatory spectroscopic observations which have yielded a further 110 new LMC PNe while confirming the 102 previously known PNe in the outer LMC. These observations, providing low and medium resolution spectra from 3650\AA~to 6900\AA,~have been added to my comparable data for PNe in the central 25deg$^{2}$ of the LMC. The combined data were used to measure fluxes in preparation for a number of projects related to luminosity functions, chemical abundances, central star properties and LMC kinematics. Here I provide a preliminary look at the range of derived central star effective temperature estimates. I also show a correlation between the central star temperatures and the expansion velocity of the nebula.
\end{abstract}

\section{Introduction}

In recent years considerable progress has been made in understanding the evolutionary sequence of planetary nebulae (PNe). The evolution of the photoionised nebula needs to be understood with regard to the processes leading to its ejection, mass/density relation, chemical composition and the post-AGB evolution of the central star. The central star in particular is the driving force, both ejecting the nebula and then releasing fast winds, driven by radiation pressure, which compress and accelerate the pre-ejected material, creating thin, ionised shells.

Since a strong link has been observationally established between the parameters of the central star and those of the surrounding nebula (eg. ~\cite{DM87, DM88},~\cite{DM90},~\cite{SVK87},~\cite{S89}), it follows that certain parameters of the central star can be determined indirectly by measuring key emission lines in the nebula. This is especially useful in the LMC where the central star cannot be directly observed.

Over the past couple of years we have used both the UKST H$\alpha$ and short red maps of the central 25deg$^{2}$ region of the LMC to uncover over 460 candidate PNe. These were labeled as `true', `likely' and `possible' depending on the quality of images and confirmatory spectra obtained. To these were added the 169 PNe that were previously catalogued in that area. Spectroscopically confirmed results including calibrated fluxes, luminosity functions and radial velocities were published in~\cite{RP06a, RP06b, RP10a}.

I have now extended our survey to the outer regions of the LMC mainly using the \OIII, \SII~and H$\alpha$~images provided by the Magellanic Cloud Emission Line Survey (MCELS). From the 1,000 or so candidates selected for spectroscopic followup, I identified 110 newly discovered and 101 previously known PNe. The complete sample, comprising 749 LMC PNe spanning the entire galaxy, has the advantage of being at a near common, known distance (49.2~kpc,~\cite{RP10a}) with low reddening, yet close enough to be studied in detail. It is currently the most complete PN sample in existence for any galaxy~\citep{R12}.

The objective of this preliminary work is to compare the temperature of the central stars to the excitation and expansion velocity of the nebulae. This allows me to investigate the evolution of both the nebula and central star as it evolves into a white dwarf.

\section{Observational data for the LMC PNe}

Follow-up spectroscopy was mainly performed on the AAT using AAOmega which comprises 400 fibres placed by robotics across a 2 degree field of view. Three nights of observations in February 2010 plus three field observations in February 2012 provided coverage of the most concentrated outer areas. For more extended outer areas of the LMC where the density of candidates was too low for AAOmega I used 6dF on the UK Schmidt telescope. This instrument operates essentially the same way but covers a larger, 6 degree area of the sky while using only 150 fibres.

Flux calibration was conducted using the method described in~\cite{RP10a} where data counts are calibrated to fluxes from HST observations for the same objects. This method has proved very reliable and allows the whole dataset to be homogeneously calibrated. Additional spectroscopy for PNe in the inner main bar regions was obtained using FLAMES on the VLT, the 1.9m telescope at the South African Astronomical Observatory and the 2.3m telescope at Siding Spring Observatory. While the long-slit spectra were reduced using standard IRAF tasks, the FLAMES multi-fibre data were flux calibrated using the method described for AAOmega and 6dF data~\citep{RP10a}.

\section{PN central star temperatures}

Without the ability to individually pinpoint and observe the central stars of LMC PNe, I use photoinisation models \citep{DJ92, RP10b} that demonstrate that for optically thick PNe in the Magellanic Clouds, the excitation class parameter is related to stellar temperature. The equation to estimate low excitation is given by:

\begin{equation}
0.45\left(\frac{F_{[OIII]\lambda5007}}{F_{\mathrm{H}\beta}}\right),~~~~~~~~~~~~~~~~~~~~~~~~~~~~~~~~~~~~~~~~~~~~~~~0.0\,<\,E\,<\,5.0
\end{equation}
while the high excitation PNe are estimated by
\begin{equation}
5.54\left[\frac{F_\mathrm{He\textrm{\sc ii}\lambda4686}}{F_{\mathrm{H}\beta}} + \log_{10}\left(\frac{F_{\mathrm{[OIII]\lambda{4959} + {5007}}}}{ F_{\mathrm{H}\beta}}\right)\right],~~~5.0\,\leq\,E\,<\,12,
\end{equation}

Using this definition, a transformation from excitation class to stellar effective temperature ($\textit{T}$$_\textmd{eff}$) was made using:

  \begin{equation}
  \textrm{log}~T_{\textmd{eff}} = 4.439 + [0.1174 \pm 0.0025] E - [0.00172 \pm 0.00037] E^{2}
    \end{equation}

    which is based on the transformation given in~\cite{DJ92} but adjusted to match the Zanstra temperatures published by~\cite{VSS03, VSS07}  (see~\cite{RP10b}). For average abundance levels within the LMC, this equation provides a useful transformation to stellar temperatures. \cite{DJ92} also expected this relation to work well, having tested it using 66 of the brightest PNe in the LMC, but predicted the relationship would break down for low excitation PNe. The reason given for this was the strong dependency of the \OIII/H$\beta$ ratio on metallicity as well as upon stellar temperature.

    In order to correct for any over-dependency on the metallicity introduced by using the \OIII/H$\beta$ ratio,~\cite{DJ92} constructed a grid, based on covering a range of stellar temperatures and metallicities using the generalised modeling code MAPPINGS \citep{BDT85}. They use an ionisation parameter defined as $\textit{Q}$ = $\textsl{N}$$_{Ly-c}$/4$\pi$$\langle$r$^{2}$$\rangle$$\textsl{N}$$_{H}$ where $\textsl{N}$$_{Ly-c}$ is the number of Lyman continuum photons emitted by the central star, $\langle$r$^{2}$$\rangle$ is the mean radius of the ionised nebula and $\textsl{N}$$_{H}$ is the nebula's hydrogen particle density. By adopting a high value for $\textit{Q}$ (2 $\times$ 10$^{8}$ cm s$^{-1}$), they simulate stellar luminosity and nebula gas pressure typical of the brighter PNe in the LMC as well as those in the Galactic Bulge. The resulting grids, encompassing abundances from 0.1 to 2.0 times solar, each with a set of temperatures between 35,000 and 140,000\,K, encompass the maximum luminosity range for PNe in both the H$\beta$ and \OIII$\lambda$5007 lines.

    Importantly, although these grids have been available for 20 years, they have not been tested against medium to faint and evolved PNe in the LMC, typical of those that would be found in the  0.0\,$<$\,E\,$<$\,5.0 excitation bracket. With our improvements to the original formulas given for excitation class and temperature, I need to investigate whether our new temperature estimates agree with the temperatures found from the modeled grid of~\cite{DJ92}.

\subsection{Results}
\begin{figure}
\begin{center}
  \includegraphics[width=1.12\textwidth]{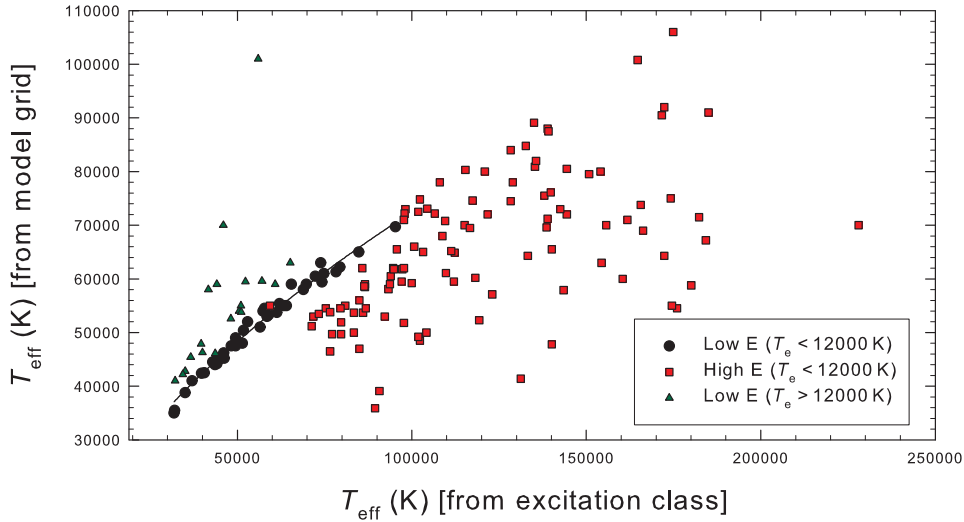}\\
  \caption{A comparison of stellar effective temperatures found from a direct reliance on excitation class and those found for the same PNe using the modeled grid of~\cite{DJ92}. Where low excitation PNe have electron temperatures below 12,000\,K there is an exponential correlation with 95\% confidence (shown curve).}
  \label{Figure1}
  \end{center}
  \end{figure}

I compared central star temperatures for high, medium and low excitation PNe, derived using our formulae (equations 1 \& 2) with central star temperatures acquired using the modeled grid of~\cite{DJ92}. The grid relies on the \OIII$\lambda$5007/H$\beta$ ratio and the electron temperature ($\textit{T}$$_\textmd{e}$) in order to produce an estimate of log (Z) and ($\textit{T}$$_\textmd{eff}$). For low excitation PNe, the similar reliance on the \OIII$\lambda$5007/H$\beta$ ratio means that the only difference will be introduced by $\textit{T}$$_\textmd{e}$. For low excitation PNe I find an exponential fit between temperatures derived directly from the excitation class (equation 3) and those derived from the grid of~\cite{DJ92}. In order to show this relation, a curve has been fitted to the data (black circles) in Figure~\ref{Figure1}. For comparison I also show the results for medium to high excitation PNe (red-filled boxes) and low excitation which do not fit the grid (green triangles).

High excitation PNe do not correlate to central star temperatures derived using excitation class (equation 2). Clearly, the reason is that high excitation PNe require the use of the HeII$\lambda$4686 line in order to obtain $\textit{T}$$_\textmd{eff}$ estimates. The \OIII$\lambda$5007/H$\beta$ ratio and $\textit{T}$$_\textmd{e}$ alone do not measure sufficient levels of excitation to permit the estimation of high central star temperatures. This result agrees with the warning given by~\cite{DJ92} in which they find that the grid is not very useful for determining stellar temperatures where $\textit{T}$$_\textmd{eff}$ $>$ 90,000\,K and log [Z] $<$ -0.5.

Although an exponential correlation is found for most low excitation PNe, there is a subgroup that return higher $\textit{T}$$_\textmd{eff}$. Using the grid, low excitation PNe with $\textit{T}$$_\textmd{e}$ higher than 12,000\,K and log (Z) less than -1.0 have increasingly higher $\textit{T}$$_\textmd{eff}$ estimates than those found using equation 1. For this reason I suggest that the grid is not useful for estimating $\textit{T}$$_\textmd{eff}$ where ($\textit{T}$$_\textmd{e}$) are greater than 12,000\,K, even though the grid allows the estimation of $\textit{T}$$_\textmd{eff}$ using $\textit{T}$$_\textmd{e}$ up to 15000\,K.

\begin{figure}
\begin{center}
  \includegraphics[width=0.79\textwidth]{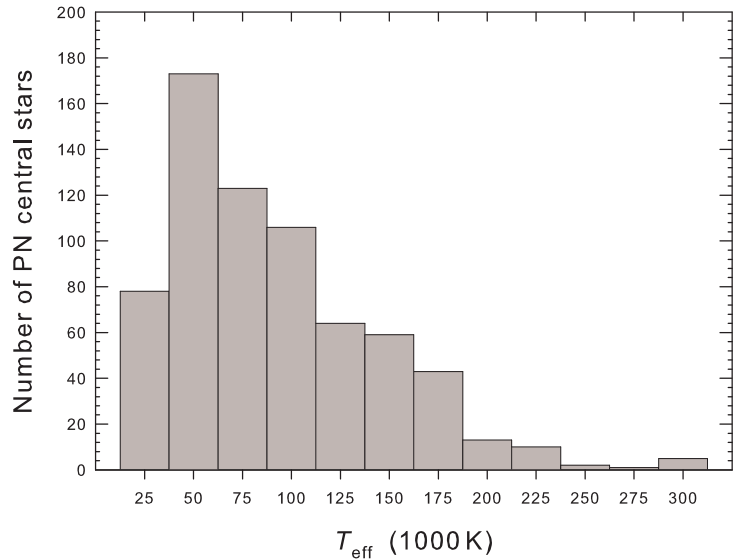}\\
  \caption{Our stellar effective temperature estimates found from a direct reliance on excitation class as derived from equations 1, 2 \& 3. The largest number of central stars fall within the 50,000\,K bin, encompassing 37,500\,K $<$ $\textit{T}$$_\textmd{eff}$ $<$ 62,500\,K. }
  \label{Figure2}
  \end{center}
  \end{figure}

The exponential curve for those low excitation PNe with $\textit{T}$$_\textmd{e}$ below 12,000\,K follows the form:

\begin{equation}
T_{\textmd{eff} [grid]} = 72.971 \times~T_{\textmd{eff} [E]}~^{0.6001}
\end{equation}

where $\textit{T}$$_{\textmd{eff} [grid]}$ is the stellar effective temperature found from the grid and $\textit{T}$$_{\textmd{eff} [E]}$ is the stellar effective temperature found from equations 1 and 3 for low excitation PNe. At low $\textit{T}$$_{\textmd{eff}}$, the grid and excitation class produce near equivalent results but as $\textit{T}$$_{\textmd{eff}}$ increases, $\textit{T}$$_\textmd{e}$ has the effect of exponentially decreasing $\textit{T}$$_{\textmd{eff}}$ estimates produced by the model. Our previous comparisons of equations 1, 2 \& 3 with $\textit{T}$$_{\textmd{eff}}$ estimates using the Zanstra method \citep{RP10b} show a good correlation where the nebulae are optically thick. In this case there is an increasing decline in grid temperature estimates where they are compared to Zanstra and excitation (equation 3) temperature estimates. Furthermore, with $\textit{T}$$_\textmd{e}$ greater than 12,000\,K the grid produces a number of the low excitation PNe with inflated $\textit{T}$$_\textmd{eff}$. This is presumably the result of an over correction for the effect of metallicity within the central star. Since there is little correlation between $\textit{T}$$_\textmd{e}$ and any method used to produce a $\textit{T}$$_\textmd{eff}$ estimate, I have decided to use equations 1, 2 \& 3 alone to estimate my central star effective temperatures for this presentation. My central star effective temperatures are shown in Figure~\ref{Figure2} where the temperatures range from 28,000\,K to 291,000\,K with a mean of 90,300\,K.

\begin{figure}
\begin{center}
  \includegraphics[width=0.82\textwidth]{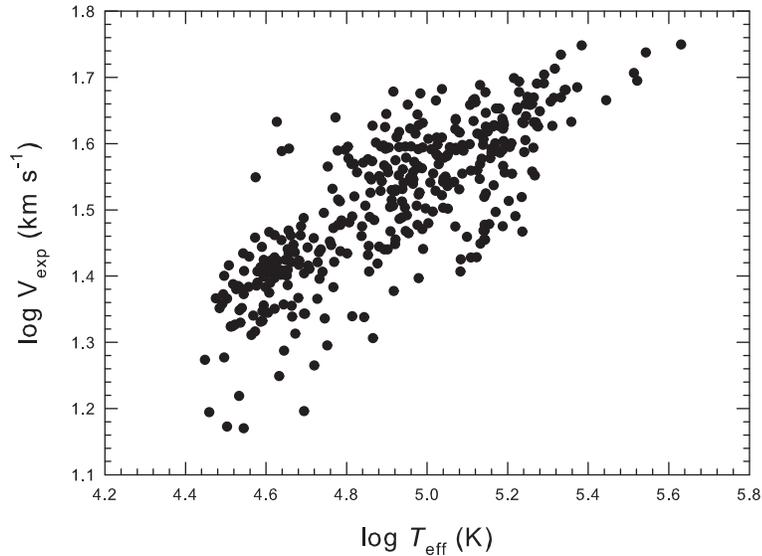}\\
 \caption{A comparison of nebula expansion velocities with stellar effective temperatures found from a direct reliance on excitation class. Points to the lower left of the plot, below the main group, are expected to be optically thin nebulae. }
 \label{Figure3}
\end{center}
 \end{figure}

  Since there is a correlation between excitation class and $\textit{T}$$_\textmd{eff}$, it follows that there is also a moderate correlation between $\textit{T}$$_\textmd{eff}$ and the expansion velocity of the surrounding nebula. In Figure~\ref{Figure3}~I show the derived expansion velocity of the nebula versus the $\textit{T}$$_\textmd{eff}$ from equation 3. This correlation was first discovered by~\cite{DF85} and later improved using a two parameter fit which included the excitation class and the H$\beta$ flux \citep{DM90}. The equation for estimating the expansion velocity is given as equation 3.2 in~\cite{DM90}.

  With a strong relationship between excitation class, the H$\beta$ flux and the Zanstra temperature of the central star~\citep{M84}, the position of a PN on plots such as Figure~\ref{Figure3}, representing the relationship between the nebula expansion velocity and $\textit{T}$$_\textmd{eff}$ will depend principally on the optical density, mass of the nebula and intrinsic properties of the central star. Since the most massive stars achieve the highest temperatures, the excitation class should also follow the mass of the star. Massive central stars fade rapidly (as seen in the brightest 4 magnitudes of the PNLF~\citep{RP10a}) so when low H$\beta$ fluxes are associated with high-excitation nebulae we can confidently assume the presence of a massive central star. Such stars drive high expansion velocities in the nebula, delivering high energy and making them more efficient at ionising the surrounding AGB wind.
 
 \bibliography{Reid_talk_a}

\end{document}